\documentclass[pdflatex,sn-mathphys-num]{sn-jnl}
\usepackage{caption}
\usepackage{graphicx}%
\usepackage{multirow}%
\usepackage{amsmath,amssymb,amsfonts}%
\usepackage{amsthm}%
\usepackage{mathrsfs}%
\renewcommand{\figurename}{Figure}
\usepackage[title]{appendix}%
\usepackage{xcolor}%
\usepackage{textcomp}%
\usepackage{manyfoot}%
\usepackage{booktabs}%
\usepackage{algorithm}%
\usepackage{algorithmicx}%
\usepackage{algpseudocode}%
\usepackage{listings}%
\usepackage{enumerate}%
\usepackage{subcaption}%
\usepackage{float}%
\usepackage{ifsym}%
\usepackage[left]{lineno}
\theoremstyle{thmstyleone}%

\theoremstyle{thmstyletwo}%

\theoremstyle{thmstylethree}%

\begin{document}
%\linenumbers
\title[Improving Spatio-temporal Gaussian Process Modeling]{Improving Spatio-temporal Gaussian Process Modeling with Vecchia Approximation: A Low-Cost Sensor-Driven Approach to Urban Environmental Monitoring}

\author*[1]{\fnm{Idir} \sur{YM}}\email{mohamed-yacine.idir@minesparis.psl.eu}

\author[2]{\fnm{Orfila} \sur{O}}\email{olivier.orfila@vedecom.fr}

\author[3]{\fnm{Chatellier} \sur{P}}\email{patrice.chatellier@univ-eiffel.fr}

\author[4]{\fnm{Judalet} \sur{V}}\email{Vincent.JUDALET@estaca.fr}

\affil*[1]{\orgdiv{Ecole des Mines de Paris}, \orgaddress{\street{35 Rue Saint-Honoré}, \city{Fontainebleau}, \postcode{77300}, \state{Ile-de-France}, \country{France}}}

\affil[2]{\orgdiv{Vedecom}, \orgaddress{\street{23 bis All. des Marronniers}, \city{Versailles}, \postcode{78000}, \state{Ile-de-France}, \country{France}}}

\affil[3]{\orgdiv{Université Gustave Eiffel}, \orgaddress{\street{5 Bd Descartes}, \city{Champs-sur-Marne}, \postcode{77420}, \state{Ile-de-France}, \country{France}}}

\affil[4]{\orgdiv{Estaca}, \orgaddress{\street{12 Avenue Paul Delouvrier}, \city{Montigny-le-Bretonneux}, \postcode{78180}, \state{Ile-de-France}, \country{France}}}

\abstract{This paper explores Vecchia likelihood approximation for modeling physical phenomena sensed by mobile and fixed low-cost sensors in urban environments. A three-level hierarchical model is proposed to simultaneously accounts for the physical process of interest and measurement errors inherent in low-cost sensors. Several innovative configurations of Vecchia's approximation are investigated, including variations in ordering strategies, distance definitions, and sensor-specific conditioning. These configurations are evaluated for approximating the likelihood of a spatio-temporal Gaussian process, using simulated data based on real mobile sensor trajectories across Nantes, France. Our findings highlight the effectiveness of the min-max distance algorithm for ordering, reaffirming existing literature. Additionally, we demonstrate the utility of a random ordering approach that doesn't require prior definition of a spatio-temporal distance. These two ordering configurations achieved, on average, 102\% better results in log Kullback-Leibler divergence compared with four other ordering schemes studied. Results are supplemented with Asymptotic Relative Efficiency analysis, offering practical recommendations for optimizing parameter estimation. The proposed model and preferred Vecchia configuration are applied to real-world air quality data collected using mobile and fixed low-cost sensors. This application underscores the model's practical value for pollution mapping and prediction in environmental monitoring. This study advances the use of Vecchia's approximation for addressing computational challenges of Gaussian models in large-scale spatio-temporal datasets from environmental monitoring with low-cost sensor networks.}


\maketitle

\section{Introduction}\label{sec:intro}

The emergence of low-cost sensors, facilitated by significant technological advances \cite{ullo2020advances}, has led to their use in real world conditions. These applications range from tracking distributed targets within autonomous systems \cite{he2019distributed}, to sampling and modeling numerous physical phenomena \cite{cheng2013latent}. Examples of such applications include sensors for measuring local weather conditions, such as pressure or temperature, low-cost sound level sensors to monitor ambient noise intensity \cite{harman2016performance}, and air quality sensors for measuring pollutant concentrations \cite{gressent2020data}.

Low-cost air quality sensors offer several advantages. Their small size and portability allow them to be used as mobile sensors, dynamically moving throughout the environment, and providing high spatial resolution data \cite{liu2012dynamic}. These sensors have the potential to bridge the data gap, influence air pollution regulations, and protect public health \cite{soppa2014respiratory}. 

However, low-cost sensors are not without shortcomings. They are subject to certain degrees of random uncertainty, including missed detection, false alarm, sensor bias, and communication-related issues \cite{nalakurthi2024challenges}. Challenges such as potential sensor drift, a common issue for low-cost air quality sensors, and potential non-linearity of sensor responses also exist \cite{giordano2021low}. Furthermore, the quality of the data from these sensors can be influenced by environmental conditions, such as relative humidity levels \cite{jayaratne2018influence}.

Despite these challenges, low-cost sensors have been widely used in various studies. For example, they have been used to generate a large volume of sampling data, allowing modeling and mapping of air pollution levels at street level resolution in urban locations \cite{hasenfratz2015deriving,chu2015modeling}. These instruments have shifted air quality monitoring from being primarily government-driven to a more commercial and/or crowd-funded approach \cite{morawska2018applications}.

Often, the phenomenon observed by these sensors is a continuous environmental process characterized by spatio-temporal dependencies. The geostatistical framework is well suited to address these challenges, offering robust methods to analyze and interpret spatial and spatio-temporal dependencies in environmental data \cite{diggle2007springer, isaacs1989applied, oliver2014tutorial, webster2007geostatistics,christakos2012modern}. These methods have naturally been applied to the problem of modeling air quality using low-cost mobile sensors \cite{idir2021mapping,gressent2020data}.

Low-cost sensors are typically characterized by limited precision in their observations. Although these sensors are usually calibrated before deployment, it is often assumed that their measurement errors are relatively small compared to the magnitude of the quantities being measured. In geostatistics, such errors are commonly modeled as Gaussian white noise, referred to as the nugget effect.

This study extends beyond the traditional approach by explicitly modeling the errors associated with low-cost sensors alongside the physical phenomenon of interest. By independently addressing the autocorrelation of observations from the same sensor, this approach provides a more realistic representation of the data, accounting for the distinct error structure introduced by the sensors.

Mixed-effects models are used for modeling autocorrelated data, assuming that observations come from different experimental units (plots, chambers, pots, etc.). One advantage of mixed-effects models is that they can be used to analyze multiple sources of variation, such as sensor effects, location effects, and time effects. These models have been applied in various environmental studies, such as air pollution modeling, and have found extensive use in social and medical sciences, biological sciences, and particularly in the field of ecology and evolution \cite{brown2021introduction}.

Gaussian processes (GP) have become a widely used tool for modeling a variety of phenomena. GPs represent a flexible framework for modeling data with arbitrary non linearities and complex dependencies, by using a specific covariance structure to capture the autocorrelation of observations, from spatial and spatio-temporal fields \cite{cressie2015statistics} to psychological models \cite{wetzels2010bayesian}.

Consequently, they find extensive application in geostatistics, machine learning, and mixed model frameworks \cite{banerjee2014hierarchical, williams2006gaussian, schulz2018tutorial}. This makes them particularly valuable, as they offer full predictive distributions rather than just point estimates, allowing for a principled approach to capturing uncertainty. The scalability of GPs represents a significant challenge, especially when dealing with large data sets that often number tens of thousands or more. Relative to the number of observations, the complexity of GP evolves quadratically for memory and cubically for time \cite{stein2004approximating}. This limits their direct application to today's large datasets, such as those derived from high-frequency sampling of low-cost sensors.
 
To address this problem, numerous approaches have been proposed in the literature. From sparse covariance matrices \cite{furrer2006covariance, kaufman2008covariance}, to sparse precision matrices \cite{rue2005gaussian, lindgren2011explicit}, and low-rank matrices \cite{higdon1998process, wikle1999dimension}. In this article, the sparse Cholesky factor of the precision matrix \cite{vecchia1988estimation,katzfuss2021general} will be used. A comprehensive review and comparison of many of these methods can be found in \cite{heaton2019case}.

\cite{guinness2018permutation} studied the impact of various observation order configurations on the quality of the Vecchia approximation in a simplified two-dimensional space, and applied it to a space time dataset. Building on this work, this study extends the evaluation of Vecchia approximation to a three-dimensional spatio-temporal space, leveraging real trajectory data from a measurement campaign with low-cost sensors. Additionally, it explores an alternative covariance function designed to account for measurement uncertainties inherent in low-quality sensors. As a result, this study will provide insights into the optimal Vecchia configuration for this specific context.

The contributions of this study include the following:
\begin{itemize}
    \item Modeling the uncertainty of low-cost sensors alongside environmental phenomena by incorporating an autocorrelation parameter specific to low-cost sensors, in addition to the traditional nugget effect.
    \item Studying Vecchia approach using a real trajectory dataset collected during a data collection campaign with low-cost sensors.
    \item Investigate various Vecchia configurations, including novel, untested setups such as spatio-temporal (ST) modeling, sensor ordering, and sensor conditioning, in the context of a more complex covariance structure.
    \item Apply the proposed approach to a set of real-world pollution data collected by low-cost sensors, comprising twenty thousands observations for mapping and prediction.
\end{itemize}

The structure of this paper is as follows. Section~\ref{sec:methods} introduces the hierarchical model and provides the motivation for its use. It also includes a brief review of the Vecchia approximation and details the configurations studied. Section~\ref{sec:results} presents the results of the performance indicators for the various configurations examined. Section~\ref{sec:application} provides a numerical analysis of the proposed model and the selected Vecchia approximation applied to real air quality data collected from low-cost mobile and fixed sensors. Finally, the article concludes with a discussion and outlines potential directions for future research.

\section{Material and methods}\label{sec:methods}

\subsection{Model formulation}

This paper aims to model air quality as measured by a fleet of low-cost mobile and fixed sensors. To achieve this, a three-level hierarchical model is proposed: one level for the unobserved latent process, another for the micro-scale variability of the phenomenon and/or observation error, and a third for the autocorrelation specific to low-cost sensors.

The modeling begins with the latent process, which represents the dispersion of pollutants in this context. Following geostatistical principles, a Gaussian process is employed in the following form:
\begin{equation}
  Y \sim N( \beta X,R(\Theta))  
\end{equation}

Where $\beta X$ represents the linear mean structure based on covariates in the design matrix $X$. A fixed mean $\beta_0$ is included by adding a column of 1s to the design matrix. $R(\Theta)$ is the covariance matrix that captures spatio-temporal dependencies, parameterized by $\Theta$. The covariance structure can be specified using kernel functions such as Matérn kernels. 

In the classical geostatistical model, the sensor error (or data model) is typically modeled by a decorrelated white noise. This means that errors are assumed to be independent and identically distributed across locations, time, and sensors. This results in the well-studied model:

\begin{equation}
     \begin{array}{l}
    Z|Y  \sim N( Y,\tau^2 I)\\
    Y \sim N( \beta X,R(\Theta))
    \end{array}
    \label{Model_01}
\end{equation}

Here, $Z$ are the data collected by the sensors, $\tau$ represents the measurement error, and $I$ is the identity matrix. This assumption may not be appropriate for data collected by multiple low-cost sensors subject to various measurement error sources that can be correlated over space and time.

The assumption of normality in this hierarchical framework serves both practical and empirical purposes. While raw environmental measurements, particularly pollutant concentrations, often exhibit strong positive skew and may be censored at detection limits, a logarithmic transformation is commonly applied to address these issues. In our application (Section~\ref{sec:application}), PM10 measurements are log-transformed prior to analysis, as pollution data typically follow a log-normal distribution \cite{hasenfratz2015deriving}. This transformation not only stabilizes variance and reduces skewness but also ensures that the Gaussian process framework is appropriate for the transformed data.

In some cases, during the deployment of mobile sensors, a specific sensor may malfunction for various reasons (technical malfunction of the sensors, blockage of the air inlet in the portable casing, etc.), resulting in errors in all the observations from one specific sensor, while the observations from other sensors remain relatively noise-free. In this case, erroneous observations from the defective sensor will result in an overestimation of $\tau$, thus reducing the weight of observations from all sensors.

To account for this kind of bias, a mixed model framework is used. This allows the model to capture the variation and bias of the errors across space-time points for each sensor, which can be affected by various factors such as sensor calibration, battery life, and environmental conditions. The mixed model framework also allows for the incorporation of additional covariates that may affect sensors, such as temperature, humidity, or sensor placement.

Observations are indexed by \( i = 1, \dots, n \), clustered into groups corresponding to sensors \( j = 1, \dots, J \). The proposed hierarchical model is described by the following system of equations:

\begin{equation}
\begin{aligned}
Z_{ij} \mid Y, a_j &\sim \mathcal{N}(a_j + Y, \tau^2I), \\
Y &\sim \mathcal{N}(\beta X, R(\Theta)), \\
a_j &\sim \mathcal{N}(0, \gamma), \quad j = 1, \dots, J.
\end{aligned}
\label{Model_02}
\end{equation}

For each sensor \( j \), the observed data \( Z_{ij} \) (where \( i \) denotes the individual observations within sensor \( j \)) follows a multivariate normal with mean \( a_j + Y \) and variance \( \tau^2 \). The sensor-specific offset \( a_j \) is assumed to follow a normal distribution with mean 0 and variance \( \gamma \), representing the bias or deviation of each sensor from the true value.

This model accounts for sensor-specific offsets (\( a_j \)). The assumption that the mean of \( a_j \) is zero implies that on average the sensors are calibrated. Assuming a nonzero mean bias would make the model non-identifiable.

Finally, as $Y$ and $a_j$ are independent, the marginal distribution of the data variable $Z$ can be obtained: 
\begin{equation}
   p(Z) = \int \int p (Z | (a_j, Y)) p (Y). p (a_j) dYda_j 
\end{equation}

resulting in a compact form of the model (\ref{Model_02}):
\begin{equation}
 Z \sim N( \beta X, \tau^2. I + R (\Theta) + K.\gamma.K' )   
\end{equation}

Where $K$ is the binary matrix $n\times J$ that codes the membership of observation $i$ to the sensor $j$, and $K'$ its transpose.

After integration, the sensor offset translates in the marginal model to a within-sensor correlation: every two observations coming from the same sensors will have an additional correlation parameter $\gamma$ in the variance-covariance matrix, regardless of their spatio-temporal distance. 

Assuming a more realistic non-zero mean bias $b$ for the offsets $a_j$, would result in the following model:
\begin{equation}
   Z \sim N (\beta X + b,  R(\Theta) +\tau^2.I + K.\gamma.K' ) 
   \label{Model_03}
\end{equation}

The model (\ref{Model_03}) will have the same likelihood for different values of the parameters of $\beta_0$ and $b$, as long as their sum is constant. This validates the definition of the statistical non-identifiability of the parameters in the model \cite{guillaume2019introductory}.

However, the model (\ref{Model_02}) is identifiable and more realistic than the classical geostatistical model. Its parameters can be estimated by maximum likelihood: 

\begin{equation}
L(\Theta, \beta, \tau, \gamma | Z) = arg\ max\  
 p(z,\Theta,\tau,\gamma) 
 \label{max_like}
\end{equation}

This model enables the prediction of both the underlying process values and the associated uncertainties at any given location and time, using the data collected from the sensors.

The proposed model can be useful in various applications such as air quality monitoring, sound monitoring, traffic surveillance, and more generally environmental monitoring, where low-cost sensors (often correlated) are used to collect data on spatial and spatio-temporal processes. 

\subsection{Vecchia approximation}

Although the estimation of the parameters of the proposed model (\ref{Model_02}) is theoretically possible by maximum likelihood, this approach becomes very challenging when dealing with large data sets, particularly when several tens of thousands of data records are reached. To address this limitation, a method is employed to approximate the likelihood function. Specifically, the Vecchia approximation likelihood is used, providing a sparse Cholesky decomposition of the precision matrix.

Vecchia approximation is based on replacing the density of model (\ref{Model_02}), which can be written as a conditional density:

\begin{equation}
 p(Z)=p(z_1)\prod_{i=2}^{n}p(z_i|z_1,....,z_{i-1})= p(z_1)\prod_{i=2}^{n}p(z_i|z_{v(i)})
 \label{true_like}
\end{equation}

The Vecchia approximation replaces the complete conditioning vectors $v(i)=1,....,i-1$ by a subvector $\Tilde{v} (i)$. This subvector is often chosen to contain indices corresponding to observations nearby in distance to the i\textsuperscript{th} observation, resulting in a valid joint distribution:

\begin{equation}
\Tilde{p}(Z)=p(z_1)\prod_{i=2}^{n} p(z_i|z_{ \Tilde{v}(i)})
 \label{approx_like}
\end{equation}

Unlike $p(Z)$, $\Tilde{p}(Z)$ now depends on the order in which the observations are classified, as well as the length of $\Tilde{v}$ (hereby denoted by $M$). 

If the entire sub-vector is used, it would revert to the original likelihood function. However, by using a smaller subvector, this allows for a significant reduction in the computational complexity of the likelihood function. The use of conditional densities also permits the parallelization of computations, making the Vecchia approximation highly efficient for large-scale problems.

In essence, Vecchia likelihood approximation provides a balance between computational tractability and accuracy, making it a powerful tool for Gaussian process modeling in spatio-temporal fields.

\subsubsection{Vecchia configurations}

The primary objective of this research is to investigate the efficiency of different configurations for a space-time Gaussian process integrated with a mixed-effects model parameter. Configurations, in this context, are strategic permutations of two crucial components: the ordering of observations and the selection of the conditioning set of observations.

Ordering of observations: in essence, ordering determines the sequence in which the data points are processed. The following six ordering configurations have been investigated:

\begin{itemize}
\item Random ordering: the observations are processed in a completely arbitrary sequence. 
\item Spatial ordering: the original ordering employed in \cite{vecchia1988estimation}, observation are ordered based on their geographic (spatial) coordinates 
\item Temporal ordering: similar to the previous order, this orders observations according to their coordinates in the temporal dimension.
\item Max-Min Distance (MMD) ordering: this method utilizes the minimum and maximum distance algorithms among the observations to create an order. The main idea is to successively choose points that are the farthest (maximizing minimum distance) from the points already preceded.
\item Middle-Out ordering: the middle-out approach begins the ordering from a central observation and moves outwards in all directions. 
\item Sensor ordering: observations are organized based on their corresponding sensors. All observations from Sensor 1 are first gathered in chronological order, followed by the observations from Sensor 2, and so on. The sensor that appears first is chosen based on the date of its earliest observation.
\end{itemize}

Conditioning set of neighbors: the conditioning set of neighbors refers to the selection of data points that are considered in calculating the conditional density for a specific data point. Generally in the literature, the $M$ nearest neighbors are chosen from those that appear earlier in the ordering. To calculate the M nearest neighbors, three distance definitions are used: 

\begin{itemize}
\item Spatial neighbors: only observations in the immediate spatial vicinity are considered. In this case, the distance is defined based solely on the spatial dimension. The reason for this choice is to encourage neighbor sets to contain observations from different time points, so that conditional likelihoods contain information about the temporal range parameter.
\item Temporal neighbors: analogous to the previous one, here only the temporal distance is used.
\item Spatio-temporal neighbors: the M nearest space-time neighbors are chosen.
\end{itemize}

In the last choice of distance, the metric spatio-temporal distance is used: 

\begin{equation}
d = \sqrt{(x_2 - x_1)^2 + (y_2 - y_1)^2 + K (t_2 - t_1)^2}
\end{equation}

The parameter K is estimated by minimizing the root mean square error, using inverse distance weighting on a subsample of the data.

In addition to the choice of the distance metric to select the nearest neighbors, $\Tilde{v}(i)$ can be varied upon the membership of the conditioning set to the sensor of observation $i$:

\begin{itemize}
\item Sensor dependent: the M observations in the conditioning set for observation $i$ are required to come from the same sensor as observation $i$, even if it means more distant observation, ensuring the presence of the parameter $\gamma$ in the conditional likelihood. 
\item Sensor independent: the set of neighbors is selected regardless of their sensor membership.
\end{itemize}

\begin{figure}[H]
\centering
\includegraphics[width=\textwidth]{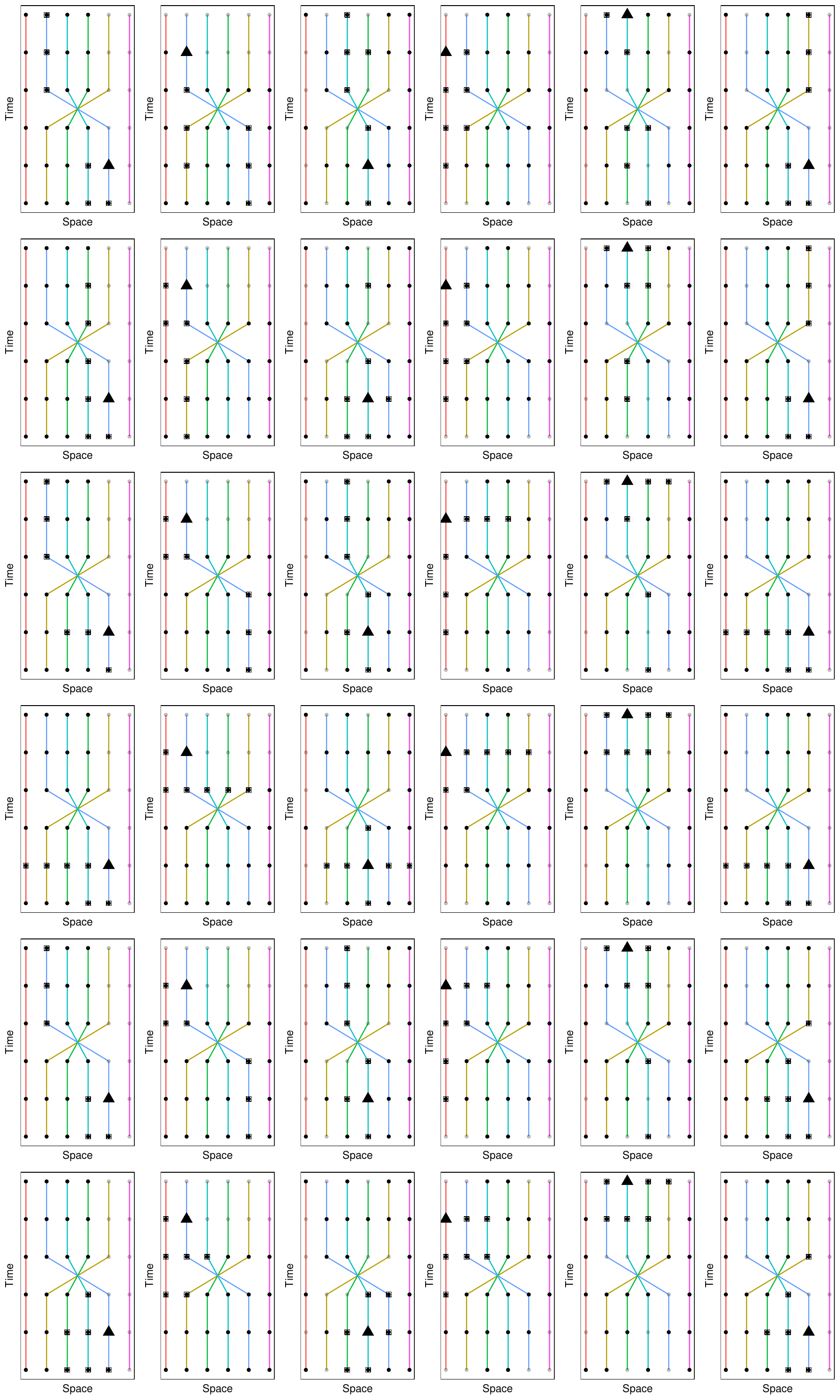}    
\caption{Configurations studied: Column indicates ordering from left to right: Spatial, Temporal, Random, Middle Out, MMD, Sensor. The first two rows indicate spatial-only distance, the third and fourth rows indicate temporal-only distance, and the last two rows indicate spatio-temporal distance. Even rows indicate conditioning primarily on the same sensors' observations. Uneven rows indicate conditions, regardless of the sensor. The black circles are ordered observation from 1 to 25, the triangle point is number 26, and the squared point represents the conditioning set of observation 26.} 
\label{fig:configurations}
\end{figure}

By cross-combining the six ordering strategies and the six neighbor selection criteria (6×3×2), a comprehensive set of configurations for the Vecchia approximation is obtained. 36 configurations are studied and evaluated to give the best approximation. For each configuration, the size M of $\Tilde{v}$ will be gradually increased, and the selected performance indicators will be computed. Figure~\ref{fig:configurations} shows a toy example in a spatio-temporal space with 36 points coming from 6 sensors.

\subsection{Performance Criteria}

The performance of each configuration is evaluated based on two primary metrics:

Kullback-Leibler (KL) divergence: This metric quantifies the difference between the approximate model derived from each configuration and the target model. The KL divergence, or relative entropy, provides a measure of the information lost when the approximate model is used instead of the true target model. A lower KL divergence indicates a closer approximation to the target model, signifying a superior configuration.

\begin{equation}
\mathrm{KL}[p \| \Tilde{p}]=\frac{1}{2}\left[\operatorname{tr}\left(\Sigma_{\Tilde{p}}^{-1} \Sigma_p\right)-\ln \frac{\left|\Sigma_p\right|}{\left|\Sigma_{\Tilde{p}}\right|}-n\right]
\end{equation}

where $\operatorname{tr}(\cdot)$ denotes the trace of a matrix , $|\cdot|$ denotes the matrix determinant, and $n$ is the number of observations. $\Sigma_p$ represents the covariance matrix of the true distribution: $\Sigma_p= R(\Theta) +\tau^2.I + K.\gamma.K'$ and $\Sigma_{\Tilde{p}}$ the covariance matrix implied by Vecchia approximation.

Asymptotic Relative Efficiency (ARE): The ARE provides a measure of the efficiency of estimating covariance parameters, computed using the generalized Fisher information for miss-specified likelihoods. The ARE for each configuration is computed as the sum of the diagonal elements of the Fisher information matrix. In essence, it captures the precision of the estimates of the covariance parameters. Lower ARE values denote more efficient estimates, thus marking a more effective configuration. 

\begin{equation}
\text{ARE}(p , \Tilde{p}) = \sum_{\theta} \lim_{n \to \infty} \frac{\text{Var}( \hat{\theta}_{\Tilde{p}})}{\text{Var}( \hat{\theta}_p)}
\end{equation}

Where $\theta$ is a parameter within the model's parameter space $\Theta$. $\hat{\theta}_p$ represents the estimated parameter in the true model, while $\hat{\theta}_{\tilde{p}}$ denotes the estimated parameter in the approximated model.

The exploration of these diverse configurations combined with the performance evaluation criteria will shed light on the optimal strategies for spatio-temporal Gaussian process modeling with a mixed effect, in the context of low-cost sensor data for environmental mapping and prediction.

\subsection{Trajectories Data and Simulation Parameters}

The cornerstone of this methodological approach to compare different Vecchia approximations with the true model lies in simulating the Gaussian process using mobile sensor trajectories. This strategy provides a controlled experimental environment in which the true underlying statistical model is perfectly known. Thereby enabling an accurate assessment of the Vecchia configurations, by allowing for the computation of performance indicators, which would not be feasible with non simulated data.

\begin{figure}[h]
\centering
\begin{subfigure}[b]{0.49\textwidth}
\centering
\includegraphics[width=\textwidth]{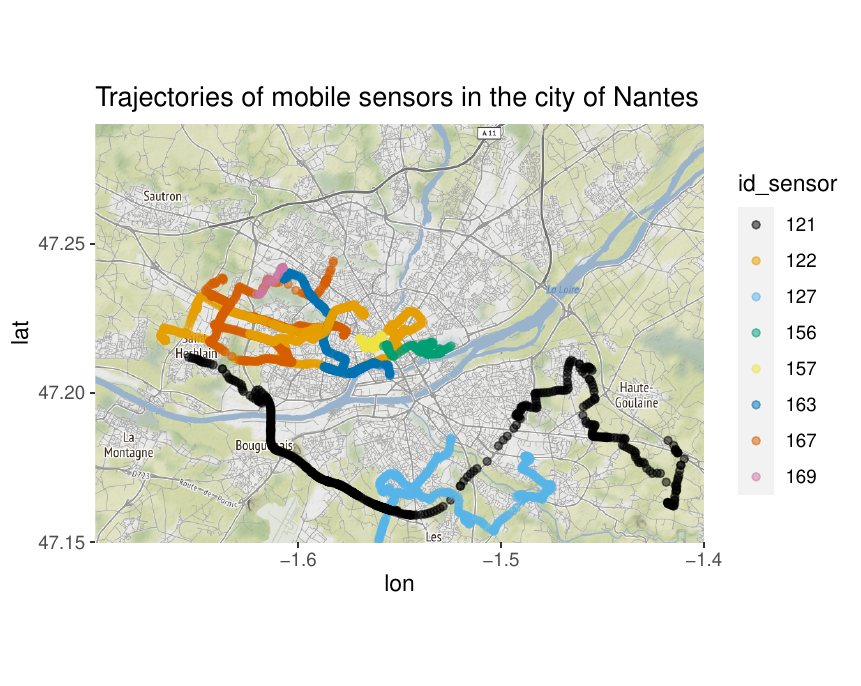}
\caption{Spatial trajectory in the city of Nantes}
\label{fig:traj_spatial}
\end{subfigure}
\hfill
\begin{subfigure}[b]{0.49\textwidth}
\centering
\includegraphics[width=\textwidth]{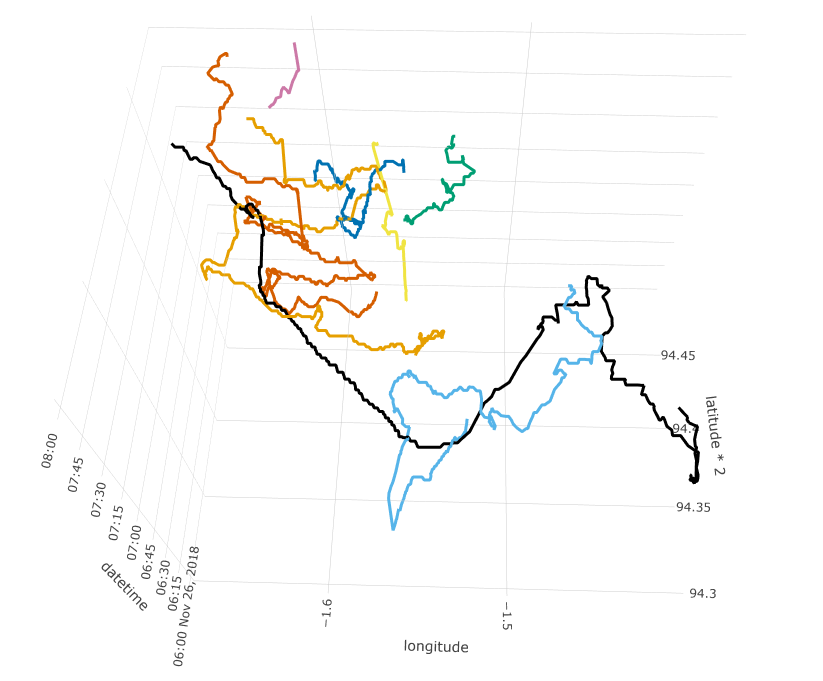}
\caption{Trajectory represented in a spatio-temporal space}
\label{fig:traj_st}
\end{subfigure}
\caption{Mobile sensor trajectories used to simulate the data set}
\label{fig:trajectories}
\end{figure}

The data set comprises simulated spatio-temporal data generated using a Gaussian process model. This simulation is based on real trajectory data collected from a fleet of mobile sensors operated by the company 'atmotrack', which criss-cross the city of Nantes, France. These sensors provide GPS coordinates and timestamps along with air quality measurements. While the measurement data are not the focus here, the spatio-temporal information forms the foundation for this simulation, ensuring that the results remain closely tied to real-world application contexts.

The Gaussian process used for the simulation assumes a mean of zero. In practice, the mean is often non-zero, but estimating and subtracting it is typically straightforward and computationally efficient. For simplicity, the mean is omitted here. The process employs an exponential metric spatio-temporal covariance function, where the covariance between two observations depends on their spatial and temporal distances, described by the formula:

\begin{equation}
\label{eq:covariancefunction}
C(\mathbf{x}_i, t_i; \mathbf{x}_j, t_j) = \sigma^2 \exp \left( -\frac{\|\mathbf{x}_i - \mathbf{x}_j\|}{\theta_1} - \frac{|t_i - t_j|}{\theta_2} \right) + \tau^2 \delta_{ij}
\end{equation}

where $\mathbf{x}$ and $t$ denote spatial and temporal coordinates, respectively, $\sigma^2$ is the process variance, $\theta_1$ is the spatial range, and $\theta_2$ is the temporal range, $\tau^2$ the nugget effect. The values for these parameters were fixed after estimation from real air quality data collected by the atmotrack sensors. This approach ensures that the simulated data set reflects the statistical characteristics of the real-world phenomena being studied.

The simulation begins by constructing a variance-covariance matrix using the spatio-temporal distances among observations. The Cholesky decomposition method is then applied to generate a set of correlated random variables. To incorporate the measurement uncertainty typical of low-cost sensors, a sensor-specific noise is added to the simulated data. This noise is drawn from a normal distribution with a mean of zero and reflects the intrinsic variability of individual sensors. The resulting data set effectively captures the spatio-temporal variability of a physical phenomenon while incorporating realistic measurement noise.

The simulated data set consists of 3000 spatio-temporal observations collected over a two-hour period from eight sensors. This combination of real-world trajectories and simulated Gaussian process values provides a robust platform for evaluating various configurations of the Vecchia approximation.

Figure~\ref{fig:trajectories} presents a visual representation of the real trajectory data, illustrating the sensor paths through Nantes and their temporal progression. This visualization underscores the connection between the real-world sensor network and the simulated spatio-temporal Gaussian process data.

\section{Comparison result}\label{sec:results}

This section presents the results of the investigation into the performance of the different configurations. Two key aspects are emphasized: the ordering of observations and the selection of the conditioning set of neighbors. The performance evaluation criteria considered were KL divergence and ARE for estimating covariance parameters.

\begin{figure}[h]
\centering
\includegraphics[width=\textwidth]{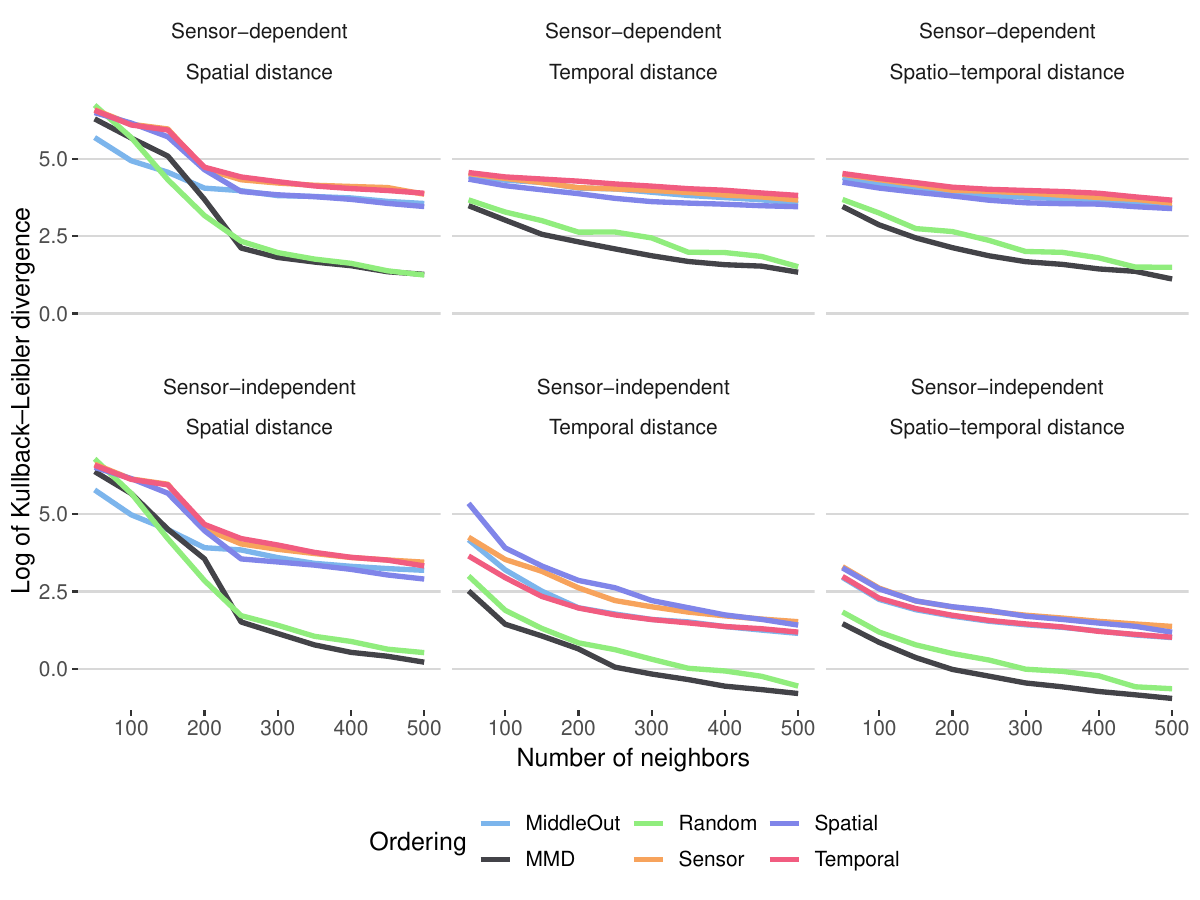}
\caption{Kullback–Leibler divergence results for the 36 configurations studied.}
\label{fig:KL}
\end{figure}

\begin{figure}[h]
\centering
\includegraphics[width=\textwidth]{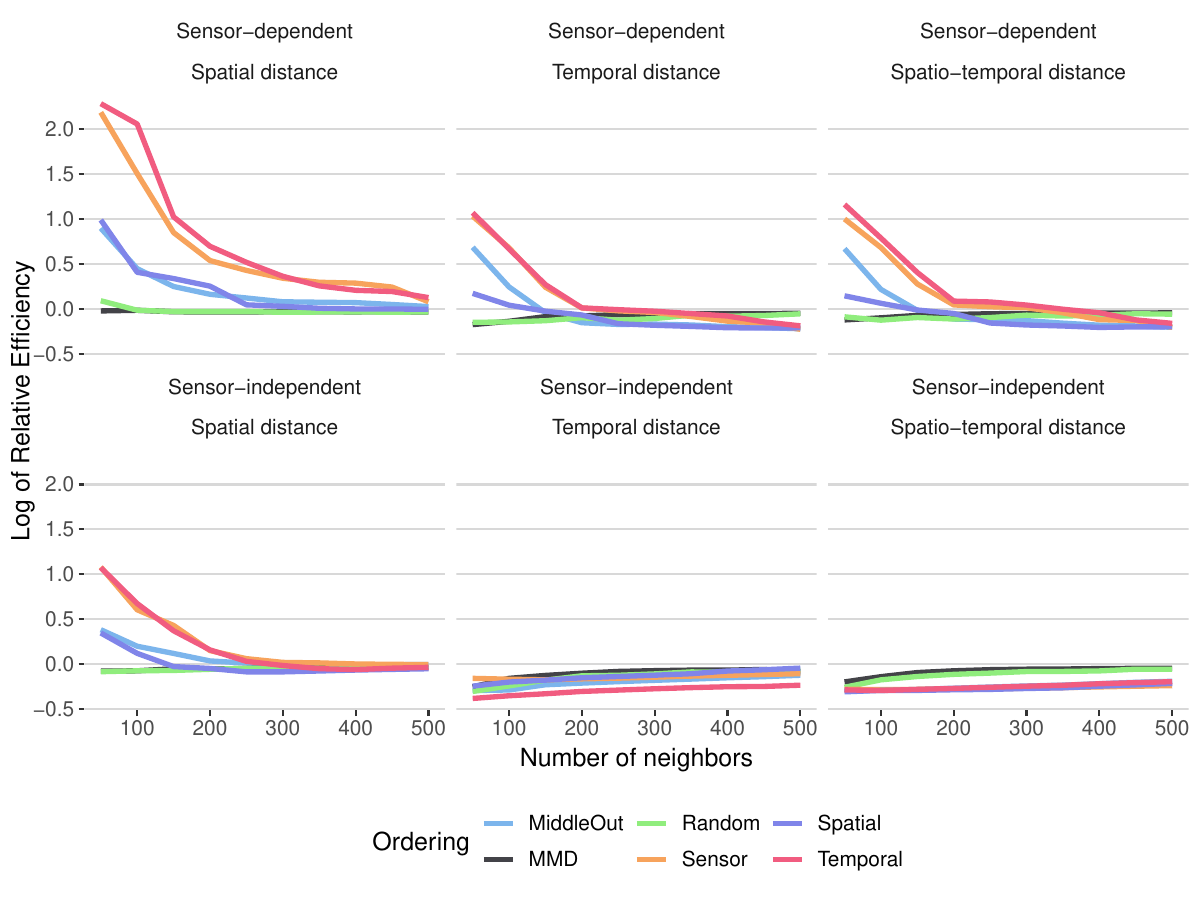}
\caption{Log of Asymptotic Relative Efficiency results for the 36 configurations studied.}
\label{fig:Efficiency}
\end{figure}

\subsection{Kullback-Leibler divergence}

The results of the KL divergence evaluation are shown in Figure~\ref{fig:KL}. Lower scores indicate a closer approximation of the true model, suggesting that these configurations capture the underlying statistical characteristics more effectively.

As expected, the divergence scores decrease as the number of neighborhoods increases, regardless of the order or the set of neighbors: the more neighbors are added, the closer we get to the true model.

The MMD order demonstrates clear advantages in the spatio-temporal context. \cite{guinness2018permutation} demonstrated its efficacy in a two-dimensional spatial setting. In the present study, it is observed that the MMD order consistently provides robust performance, and the random order also produces highly competitive results, extending \cite{guinness2018permutation} findings in this present context. In certain instances, the random order even surpasses the MMD order in terms of performance.

Conditioning solely on observations from the same sensors is suboptimal. The configuration where no such sensor constraints are applied yields better results across all orders, as it facilitates the establishment of closer neighborhoods.

The choice of distance metric is critical, particularly for small values of $M$. Spatial distance alone performs significantly worse than temporal and spatio-temporal distances when $M$ is small, although the difference decreases as $M$ increases. Overall, the spatio-temporal distance is preferred, regardless of the number of neighborhoods used. However, its use requires estimating an additional parameter prior to analysis. If avoiding this step is preferred, purely temporal distance remains a highly effective alternative.

The observed asymmetry between the effects of spatial and temporal distances likely arises from the scale at which the data are analyzed. The preference for temporal distance remains valid in contexts where measurements are collected at a city scale, with high-frequency data sampled every second over a few hours. In such cases, the use of temporal distances yield better approximations.

\subsection{Asymptotic Relative Efficiency}

The results of ARE are shown in Figure~\ref{fig:Efficiency}, with the results presented on a logarithmic scale to facilitate analysis. A score of 0 indicates that the ratio between the variance of the estimated parameters in the approximation and the variance of the parameters of the true model is equal to 1.

As observed with the KL divergence, the MMD and random orders exhibit the closest alignment with the true model, resulting in relative efficiencies closer to 1. Additionally, increasing the number of neighbors consistently drives the parameter variance toward that of the true model, across all orders and neighborhood configurations.

Interestingly, in specific scenarios, the ARE can drop below 1, indicating that the variance of the estimated parameters in the approximate model is lower than that of the true model. This occurs in configurations where neighborhoods are not constrained by sensor origin and where temporal or spatio-temporal distances are used. Such results highlight the advantage of avoiding sensor-specific constraints: not only are the approximate models closer to the true model, but the parameter estimates also exhibit reduced variance.

For configurations employing temporal or spatio-temporal distances without sensor constraints, the temporal order consistently results in smaller parameter variance. This variance gradually increases with $M$ until it matches the variance of the true model.

The KL divergence is retained as the primary evaluation metric. Notably, a configuration that performs poorly in KL divergence but demonstrates strong performance in ARE suggests improved parameter estimation for an offset model, a scenario of limited practical interest.

In conclusion, configurations without sensor-based restrictions, particularly those utilizing temporal or spatio-temporal distances, are superior for parameter inference. While spatio-temporal distances require estimating an additional parameter to define an optimal metric, the improved accuracy alone does not always justify this extra step. Regarding the choice of order, the random order is advantageous as it does not require the estimation of distance parameters, whereas the MMD order consistently yields excellent results. If minimizing parameter variance is a priority, alternatives such as the temporal order may also be considered.

From a computational perspective, the random ordering offers additional advantages beyond approximation quality. Unlike MMD ordering, which requires computing all pairwise distances to determine the ordering sequence, random ordering eliminates this preprocessing step entirely. For large datasets with $n$ observations, the MMD ordering requires $O(n^2)$ distance computations, whereas random ordering has effectively zero ordering cost. This computational saving becomes increasingly significant as dataset size grows, making random ordering particularly attractive for real-time or near-real-time applications where computational efficiency is critical.
 
\section{Numerical application}\label{sec:application}

This section presents the numerical application of the optimal configuration (maximum minimum distance order with spatio-temporal distance) to estimate the parameters of the Vecchia likelihood of the model (\ref{Model_02}) described in Section~\ref{sec:methods}. The application uses the air quality dataset collected by a fleet of fixed and mobile sensors provided by Atmotrack in Nantes. First, a brief introduction to the data is presented, followed by the results of the mapping process results.

\subsection{Air quality data}

Atmotrack (atmotrack.fr) is a company based in Nantes, France, specialized in air quality measurement. It operates a fleet of low-cost mobile sensors placed on utility vehicles such as those used by the post office, as well as fixed low-cost sensors placed in strategic locations in the city of Nantes, notably near fixed governmental air quality stations.

The data set consists of air quality measurements of PM2.5 and PM10 collected throughout November 2018. For this analysis, only data from a single day, November 26, was considered, as it contains the most comprehensive observations. The study area is illustrated in Figure~\ref{fig:trajectories}.

The retained data set includes a total of 19,523 observations, with 14,078 collected from 9 mobile low-cost sensors and 5,445 from 7 fixed low-cost sensors. The PM10 measurements were used for this study.

The low-cost sensors were pre-calibrated with the fixed stations by Atmotrack. To maintain the integrity of the data, a minor post-processing procedure was carried out. Specifically, the first five minutes of mobile runs were excluded from the data set, as it was observed that during this period the mobile sensors produced distorted data. Following this, a running median of 15 observations per sensor was applied over time to smooth out sensor errors. More information on the Atmotrack data set can be found in \cite{gressent2020data}.

The pollution data from the low cost sensors were transformed into a logarithmic scale. The literature has shown that pollution measurements often follow a log-normal distribution \cite{hasenfratz2015deriving}. 

In addition to the data from Atmotrack's low-cost sensors, representing the variable of interest $Z$, 13 additional covariates were used, primarily sourced from Open Street Map and the chemistry-transport model CHIMERE. These covariates provide information about the environment surrounding the observations. The study incorporated purely spatial, purely temporal, and spatio-temporal covariates.

Different studies have used a variety of categories of covariates, such as data from fixed monitoring stations in \cite{idir2021mapping}, annual average pollution data in \cite{gressent2020data}, or spatial land use covariates in \cite{hasenfratz2015deriving}. These covariates will be used to estimate the mean in model~(\ref{Model_02}).

\begin{table}[h]
\caption{Summary of covariates used in this study.}\label{tab:covariate}
\begin{tabular}{@{}llll@{}}
\toprule
Variable & Origin & Dimension & Temporal Resolution\\
\midrule
Temperature & public.opendatasoft & Temporal & 1 hour\\
Humidity & public.opendatasoft & Temporal & 1 hour\\
Elevation & data.nantesmetropole & Spatial & Static\\
Proximity to: motorway & Open Street Map & Spatial & Static\\
Proximity to: trunk & Open Street Map & Spatial & Static\\
Proximity to: primary & Open Street Map & Spatial & Static\\
Proximity to: secondary & Open Street Map & Spatial & Static\\
Proximity to: tertiary & Open Street Map & Spatial & Static\\
Proximity to: residential & Open Street Map & Spatial & Static\\
Proximity to: green spaces & Open Street Map & Spatial & Static\\
Proximity to: industrial site & Open Street Map & Spatial & Static\\
Nb of buildings in 500m & Open Street Map & Spatial & Static\\
Proximity to: river & Open Street Map & Spatial & Static\\
CHIMERE analysis & CAMS & Spatio-temporal & 15 minutes\\
CHIMERE forecast & CAMS & Spatio-temporal & 15 minutes\\
\botrule
\end{tabular}
\end{table}

To ensure that each low-cost sensor observation is associated with the corresponding covariates, interpolation is necessary, as some low-cost sensors sample data as frequently as one observation per second, while certain covariates are only available at intervals of 15 minutes or one hour. Table~\ref{tab:covariate} provides a summary of all the covariates used in this study, including their source and dimension.

Parameter inference for model~(\ref{Model_02}) was conducted using the exponential metric function as the covariance function~(\ref{eq:covariancefunction}), with a linear combination of covariates as mean. The Vecchia approximation was employed to maximize the likelihood, using the MMD order and spatio-temporal distance without imposing constraints on sensor membership.

\subsection{Results}

The results of parameters estimation for models~(\ref{Model_01}) and (\ref{Model_02}) are presented in Table~\ref{tab:estimation}. As anticipated, model~(\ref{Model_02}) exhibits a higher likelihood, as model~(\ref{Model_01}) is nested within model~(\ref{Model_02}). This nesting ensures that the likelihood of model~(\ref{Model_02}) is at least equal to that of model~(\ref{Model_01}) when $\gamma = 0$. In the presence of the additional parameter $\gamma$, an estimated value of 0 for $\gamma$ would result in a reversion to model~(\ref{Model_01}), yielding the same likelihood.

Furthermore, a significant portion of the variance in the model~(\ref{Model_01}) has been captured by $\gamma$.

\begin{table}[h]
\caption{Result of parameter estimation procedure.}\label{tab:estimation}
\begin{tabular}{@{}lll@{}}
\toprule
Parameter & Classical model 1 & Hierarchical model 2 \\
\midrule
Variance $\sigma^2$ (log($\mu g/m^3$)$^2$) & $1.9 \times 10^{-1}$ & $2.6 \times 10^{-2}$ \\
Spatial range $\theta_1$ (meters) & $1.38 \times 10^{4}$ & $4 \times 10^{3}$ \\
Temporal range $\theta_2$ (seconds) & $8.1 \times 10^{3}$ & $10^{3}$ \\
Nugget $\tau^2$ (log($\mu g/m^3$)$^2$) & $5 \times 10^{-4}$ & $2.9 \times 10^{-2}$ \\
Sensor autocorrelation $\gamma$ (log($\mu g/m^3$)$^2$) & -- & $2.6 \times 10^{-2}$\\
\midrule
Log-likelihood & 35112 & 37924 \\
\botrule
\end{tabular}
\end{table}

After parameter inference, any new prediction at an unsampled location $Z_0$ results in the vector $(Z_0, Z)$ following a multivariate Gaussian distribution. Consequently, the conditional expectation is used to generate a spatial interpolation map. The interpolations were computed on an evenly spaced grid of size 100×100 in latitude and longitude across twelve time periods, from 7 a.m. to 6 p.m. resulting in 10,000 points per hour and covering a total of 100×100×12 points. This grid encompasses the entire city of Nantes and spans the active daytime hours.

As new observations require corresponding covariates, those derived from Open Street Map were computed directly, while the remaining covariates were interpolated.

\begin{figure}[H]
\centering
\includegraphics[width=\textwidth]{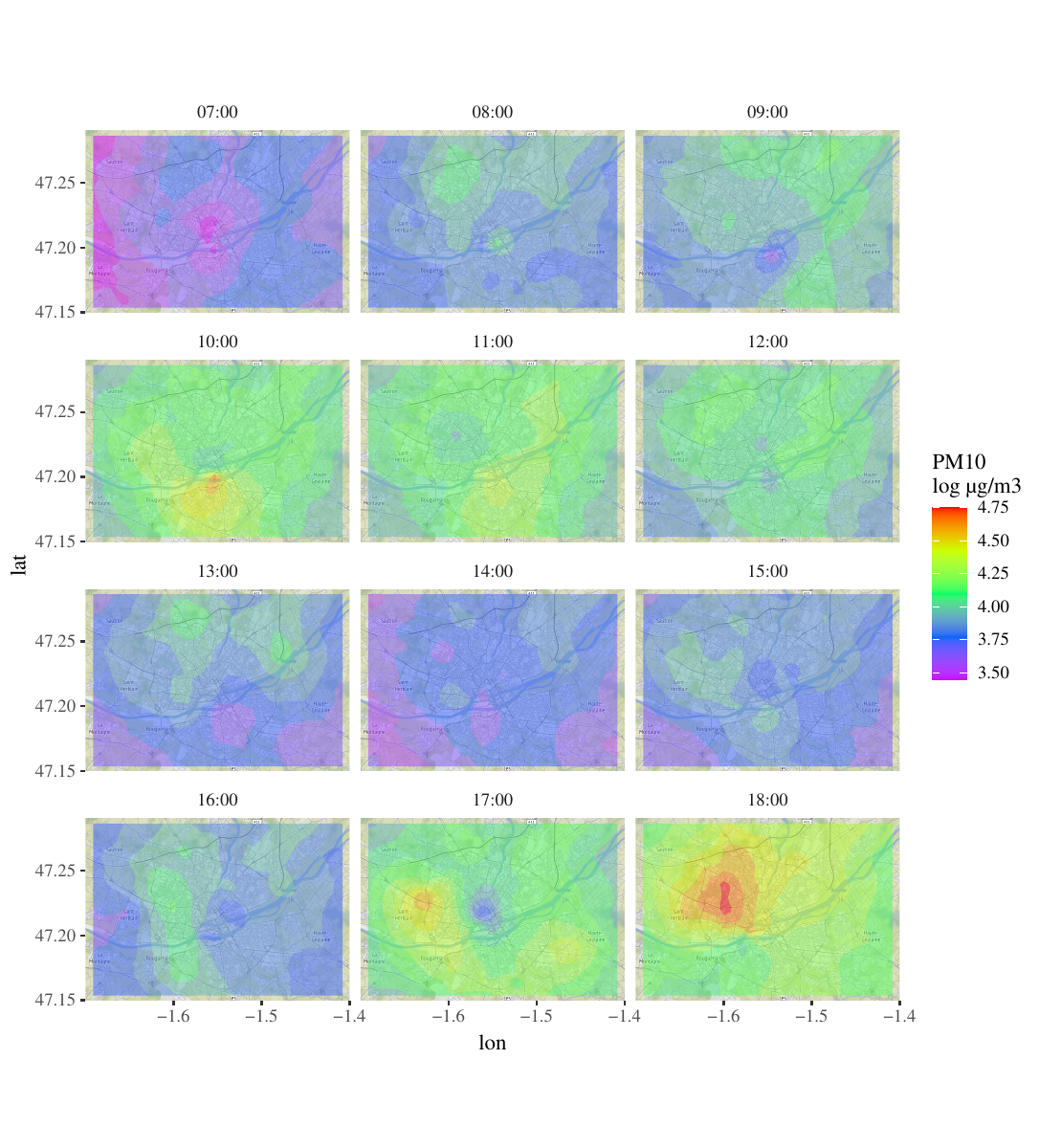}    
\caption{Mapping resulting from the application of the proposed model using the Vecchia approximation} 
\label{fig:map}
\end{figure}

Figure~\ref{fig:map} demonstrates the successful application of this methodology to the data set presented in the preceding subsection. Daily peaks were detected at 9-10 a.m. and 5-6 p.m., reflecting the average trend driven by road traffic during rush hours. As shown in the panel for 6 p.m., the methodology allows detection of the neighborhood with the pollution peak, in contrast to the global data provided by traditional information services, which offer a city-wide average. The use of the Vecchia approximation enables fast, near real time predictions and mapping of air quality, offering valuable insights to inform the public for better decision-making and enhanced environmental awareness.

\section{Discussion}\label{sec:discussion}
The results of relative efficiency (Figure~\ref{fig:Efficiency}) encompass the variances of all parameters. However, when focusing on a single parameter, such as $\gamma$, the relative efficiency of that particular parameter may be better in a specific configuration. When considering all parameters simultaneously, the relative efficiency graph provides guidance for identifying the optimal configuration.

While KL divergence was retained as the primary evaluation metric for selecting the best Vecchia approximation configuration (as it measures overall model fidelity), the ARE results presented in Figure~\ref{fig:Efficiency} provide complementary insights that are crucial for parameter inference. There are cases when it may be more beneficial to select a configuration with slightly higher KL divergence in order to obtain better estimation of the parameters, particularly when the primary goal is accurate parameter inference rather than prediction. For example, if precise estimation of the spatial or temporal range parameters is critical for scientific interpretation, the temporal order could be chosen as a determining factor in such scenarios, despite not having the absolute lowest KL divergence. This trade-off between model approximation quality and parameter estimation efficiency highlights the importance of considering both metrics when selecting a Vecchia configuration for a specific application.

In addition to providing parameter variances, the Fisher information matrix also offers insight into the covariance between parameters. Although not shown in the results section, a significant covariance was observed between the parameters $\sigma$ and $\gamma$. This covariance arises from the model formulation and the fact that most observations in the conditioning set originate from the same sensor due to spatial and temporal proximity.

Alternative conditioning configurations were investigated but not displayed, including conditioning solely on observations from sensors other than the sensor of observation $i$. As expected, the results were even worse, as this typically conditions on more distant observations.

\section{Conclusion}\label{sec:conclusion}

In conclusion, this study has demonstrated the effective application of Vecchia approximation in a spatio-temporal Gaussian process with a mixed-effects parameter, along with the optimal configurations for its implementation. The model was applied to data collected from low-cost mobile and fixed sensors in the city of Nantes, offering a practical and real-world context for the findings.

The results revealed that both the random ordering and the minimum maximum distance (MMD) provided the most accurate results for Vecchia approximation. This finding is significant as it offers a robust method to handle the inherent error or correlation in low-cost sensor measurements.

The successful application of this model to a data set of air quality measurements underscores its potential for a wider application in the field of environmental monitoring. Low-cost sensors are becoming increasingly prevalent due to their affordability and portability, and this study provides a promising direction for future research in this area.

Further studies could explore the application of this model to other types of spatio-temporal physical phenomena and in different urban environments. This would further validate the findings presented in this article and broaden the applicability of the model.

\backmatter

\bmhead{Acknowledgements}

This research was supported by ESTACA. The authors declare no conflicts of interest related to this work. We would like to express our sincere gratitude to Atmotrack and Valentin Gaufre for generously providing the dataset used in this study.

This study was conducted using R \cite{R-base}, with modeling facilitated by the ``GpGp'' package \cite{GpGp}.

\section*{Funding}
This research was funded by ESTACA.

\section*{Declarations}

\begin{itemize}
\item \textbf{Funding:} This research was supported by ESTACA.
\item \textbf{Conflict of interest:} The authors declare no conflicts of interest related to this work.
\item \textbf{Data availability:} Data available upon request from Atmotrack.
\item \textbf{Code availability:} Code available upon request from the corresponding author.
\item \textbf{Author contribution:} All authors contributed equally to the conception, design, analysis, and writing of this manuscript.
\item \textbf{Ethical responsibilities:} All authors have read, understood, and have complied as applicable with the statement on "Ethical responsibilities of Authors" as found in the Instructions for Authors.
\end{itemize}
\bibliography{bibliography}

\end{document}